\documentclass{article}
\usepackage{graphicx}

\begin{document}
\title{The Problem of Quantum Measurement and Entangled States: Interactions Among States and Formation of Detector Images}
\author{Fariel Shafee\footnote{fshafee@alum.mit.edu.  This paper contains reviews from several papers and also an unpublished arxiv draft created while the author was at the Dept of Physics, Princeton University, Princeton, NJ.  She has now left Princeton.}}
\date{}
\maketitle
\begin{abstract}
We first review and critically examine some basic concepts and ambiguities related to quantum mechanics and quantum measurement to understand the success and shortcomings of current theories. We also touch on ideas regarding expression of variables within a complex system. Then we discuss a model for quantum measurement proposed by us, in which the quantum system is allowed to interact with its image created within the detector, followed by a first passage random walk in the Hilbert space.  Definitions and ideas from the first part are used in the context of the model.  In the end, we discuss the puzzling question of entanglement.  We propose how borrowing concepts from our model for quantum measurement may enable us to formulate a hidden variable scenario that does not violate Bell's inequality.
\end{abstract}

\section{Introduction}
Recently \cite{Shafeewalk, Shafeewalkarx} a model for quantum measurement was proposed by the author.  Later \cite{Shafeebell} a draft was written to describe a situation using concepts of interaction among variables by broadening the idea of variables in the context of Bell's inequality.  Here, the quantum measurement scheme proposed by us is reviewed and the draft \cite{Shafeebell} is extended into a paper form.  The main theme of this paper is the idea of image formation within a detector, based on interactions among variables of a quantum system and the detector.

The paper is divided into three parts.  The first part reviews and analyzes some concepts and historical developments related to quantum mechanics and quantum measurement that will be valuable in the development of our model. The critical analysis of ideas will point to the gaps and inconsistencies in current theories.  In this part, we also review some concepts regarding complex systems and expression of states within complex systems. These ideas will become important when examining the interface between the quantum and the classical world, and the expression of the quantum state within the classical world.
The second parts reviews the model for quantum measurement proposed by us \cite{Shafeewalk}.
The third part uses concepts clarified in the first two parts to explore the fascinating phenomenon of quantum entanglement \cite{AC1} and the measurement of an entangled pair.

\part{Review of Existing Concepts and Some Ambiguities}
\section{Wave and Particle in Quantum Theory}
One of the main tenets of quantum mechanics is the Heisenberg uncertainty relation that states that the product of the uncertainties of the position and the momentum of a particle is related with the Heisenberg constant, h \cite{BU1}, making it impossible to know both the position and the momentum of a quantum particle simultaneously.  This notion of uncertainty leads to wave particle duality, in which particle and wave natures are interchangeable. This is manifested when electrons allowed to escape through two closely placed small parallel holes are seen to produce interference patters characteristic of waves on a screen held in front of the holes \cite{slit}. It has been argued that after leaving the holes, the electrons exhibit wave properties, and these waves are later collapsed to particles when the electron are detected.  Although, after an electron is allowed to pass through the holes, a single electron is detected in the end as well, the position of the electron is certain only upto the allowable regions of the wave-function.  It was assumed that the wave had again collapses into an electron at a specific position upon reaching the detecting screen.

This eventually gave rise to the Schrodinger formulation of quantum mechanics, where the time evolution of the wave is determined by its energy, and is given by the equation
\begin{equation}
- \frac{h^2}{4\pi m}\frac{\partial^2\Psi(x,t)}{\partial x^2}+U(x)\Psi(x,t)=i\frac{h}{2\pi}\frac{\partial\Psi(x,t)}{\partial t}
\end{equation}

 In quantum mechanics, not only the position, but certain quantum properties are uncertain to the classical world, until they are measured. An operator acting on the wave yields the set of eigenvectors, each expressing a measurable state with an eigenvalue corresponding to the value of the observable.  According to the Copenhagen interpretation, each measurement collapses the quantum wave into one of the possible quantum states.  In the quantum world, these states coexist as superposed waves but in the classical world, only one of many possible orthogonal states is expressed when measured.  This process of obtaining one of the possible quantum states in the classical world is called the measurement problem, and is one of the fundamental unsolved problems of quantum mechanics.

 Many proposals have been brought forth to explain the interface of a wave-like quantum world and a defined classical world. It has also been suggested that particles always exist, but the quantum wave nature is expressed as a limitation of the classical world. These theories proposed to validate Einstein's argument that ``God does not play dice,'' include the Pilot Wave Theory, \cite{broglie} and the Bohm Interpretation of Quantum Mechanics \cite{bohm}.  On the other hand, models that attempt to explain collapse of quantum waves as a process of spontaneous reduction, by means of modifying Schrodinger's equation or adding terms to it, have also been proposed \cite{DI1}.  Decoherence is another concept coined to explain the suppression of superposition and hence the loss of interference terms seen in the quantum world \cite{SC1,ZUREK}.  In the process of decoherence, quantum states get entangled with specific environment states and hence get separated from one another since the environment states corresponding to the different quantum states are orthogonal because of the many degrees of freedom characteristic of the environment.

 Although all these models, to some extent, succeed in explaining aspects of the process of measurement, none is complete, and many contain unexplained terms and phenomena.  For example, it is unclear how the spontaneous collapse takes place in the models of reduction, and in decoherence, it is not clear what happens to each of the decohered quantum-environment orthogonal pairs.

  The concept of diffusion with respect to quantum mechanics is not new \cite{NAS}. A relatively new model \cite{omnes} has attempted to propose reduction as the next step of decoehrence by using random walk equations to reduce the decohered states.  However the fate of the reduced states remained unclear.  Our model \cite{Shafeewalk, Shafeeimage} addressed these issues by proposing interactions among variable-states in the quantum system and within the detector. An image of the quantum system is formed within the detector, and then each quantum state and the corresponding image state form a pair.  The pairs are then subject to first passage random walk in the Hilbert space.  At the end of the random walk, one of the pairs gets amplified within the detector while the others disperse into the environment.  Notions from this model for quantum measurement proposed by us \cite{Shafeewalk} are then used later in the paper in context with the entanglement problem.

However, we first discuss some basic concepts and definitions which we will be using in our models.

\section{Transition from Quantum to Classical}
Electrons and neutrons exhibit quantum properties in double slit experiments \cite{slit} whereas tennis balls do not.  Several models try to explain why a macroscopic object does not display quantum superpositions.

Macroscopic objects such as balls themselves consist of smaller components and hence should have quantum wave properties associated with each of the components. It can be shown that after decoherence, the cancelation of random phases of many quantum waves held together result in a well defined and sharp function giving rise to the specific object \cite{ZUREK1}.

The probability of the perpetuation of rigid bodies as macroscopic objects while allowing the quantum world to behave according to the quantum Hamiltonian subject to stochastic localizations within the same modified Shrodinger equation was explored \cite{GH1}.  Again, models have been proposed that include damping terms with gravity that break down quantum equations for massive classical bodies \cite{DI1,DI2}.  In these latter models  the schrodinger equation has a gravity dependent term added to form a Shcrodinger-Newton equation, so that a massive classical object is localized with a rigorously fixed center of mass.  Alternatively a second modified equation involving the Von Newman equation used a damping term with gravity that causes decay of spatially separated superposed states in a massive classical object. The models make use of differences in mass densities of two classical state configurations.  These equations together may explain why macro objects cannot have superposed states.

However, although the lack of quantum superposition in final macroscopic states can be explained by these models, the interface between the quantum and the classical worlds, which makes the classical world see or measure a quantum object in a reduced state has not been explained satisfactorily.

\subsection{Uncertainties and Changes}
 An electron exists in a spin up and a spin down state concurrently in the quantum world, but there is no superposition of a dead cat and live cat. Uncertainties of states in a quantum world are seen to reduce to definite states in the classical world.  However, in the classical world too the states change with time and histories can be explained as continuous or sudden changes of states of macroscopic objects in the form of reorganization or redistribution of components.  A live cat can change into a dead cat by dying.  It has been suggested that the emergence of the arrow of time is specific to the process of decoherence \cite{ZUREK3}. The crux of understanding the process of quantum measurement is understanding what the states are in the quantum and the classical world, how they are connected and how they interact among themselves within the different hierarchy levels of a complex system to express the existence of one object onto another given the in-built inaccuracies and couplings at the interfaces of communication/interaction.

\section{Ambiguities of Quantum and Classical}

The ambiguous quantum states that can coexist as superposed waves represent some very specific properties.  The most basic uncertainty is the Heisenberg uncertainty relation that relate an uncertainty of momentum with the uncertainty of position in the measurement of a quantum micro particle.  Other such quantum properties, represented by the observables of quantum operators, include spin. When an electron is measured in the classical world, it can come in a spin up or a spin down state.  The x, y and z components of the spins are non-commuting, meaning the measurement of the spin of a quantum entity along one of the components destroys any previously measured certainty of the spin along the other components.  Hence, a quantum particle cannot be described to be simultaneously in a specific spin x, spin y and spin z state just like it cannot be described at a specific momentum and position state simultaneously. Characteristics like mass and charge do not exhibit such uncertainties.  An electron always has a specific rest mass when measured.

However, an object of the classical world can be described simultaneously with a momentum and position. The object itself may undergo changes and may transform into another object, with a different momentum and position when it interacts with other objects.  The laws of physics can be used to more or less predict a history of the evolution (disregarding the many degrees of freedom and information and coarse graining).  This is different from the measurement of a quantum object, which does not show a definite history of being in a spin up state or in a spin down state in the quantum world prior to measurement unless the spin in the same axis was measured before, and that value has been collapsed (so that the spin was not measured along another axis prior to the two consecutive measurements along the same axis). 

A closed quantum system preserves the uncertainties among non-commuting variables. It is possible to bring back such systems to a previous state, since operators preserving the quantum properties are unitary and hence inverse operations can lead to the initial state being repeated.  However, the many degrees of freedom in the classical world gives rise to an arrow of time in the direction of increasing entropy within a closed classical system.

The expression of a quantum characteristic such as spin in the classical world is due to the reorganization and transformation of objects or states in the classical world, that are governed by the laws of physics.  For example, the macroscopic reorganization of the detector components, changing the velocity of the pointer to make it move and point to a certain direction makes the classical world aware of a certain value of the spin. The constituents of the pointers move to the new position together and the entire pointer changes its position as the environment reorganizes to new states. Clusters of matter move together, and split or reorganize in bundled forms to make the new states emerge.
Each time the pointer moves towards a certain direction, a specific spin state is indicated.  However, the macroscopic world the detector is attached to, as a whole, evolves towards a different, higher entropy state.  Hence, although the detector may indicate the same state after different measurements, the entire environment does not remain the same during consecutive measurements.  Even the detector evolves as it ages, though certain properties of the detector are identifiable, and are expressed as a state of the detector that indicate a certain spin within the environment.
Hence, the notion of a state within the classical environment may consist of a bundled of preserved characteristics \cite{Shafeeentropy}.

Conservation of mass, charge and momentum in the macroscopic world produce consistent laws of physics that can predict results of interaction among macro objects. Let us define quantum mechanics with waves in a manner so that properties like mass and charge are distributed throughout the waves, and let us take the classical world as an approximation of a large complex quantum wave function displaying macroscopic states.  Any splitting of these states, making matter exist in more than one clustered form or organizations together as superposition would imply either creating more matter, and hence mass, or splitting mass components to exist at two places, hence reducing the mass of each state. However, since these massive particles are held together by gravity and other forces, and within the environment where these objects reside, the same building blocks such as electrons and protons etc make up the bulk of organized bodies, such a split would cause matter to reduce in mass etc in an object each time a measurement is carried out unless the exact same object can be recreated after a measurement is taken. Dissipation of energy occurring in the process of transformation of these classical states make it impossible for the world to return to the same state altogether after an event. A splitting of the worlds in the manner described by Everett \cite{everett}, might also imply a reduction of mass of each of the worlds (since in quantum mechanics the mass of an object is spread across the wave).  When two worlds are split as two quantum states, they would have different macroscopic sate organizations, and hence different futures, because of the existence of the laws of physics existing separately in each.

\section{Expression of States and Interaction}
\subsection{Variables and States}
A variable is a measurable quantity that can exist in one of many possible states. The concept of variables and states is used extensively in statistical mechanics.  In complexity theory too, the idea of variables and states is crucial.  The expression of variables within complex systems have been studied previously \cite{Shafeesoc}. The role of variables expressed at each level in a hierarchical structure has been examined \cite{Shafeehier}.  The notion of leaking information among levels in a hierarchical system because of interactions has also been suggested \cite{Shafeehier} so that the components of one level is expressed at another level of a stratified structure when information is allowed to leak. The leakage of information among variable-containing registrars or cells may also give rise to new identities and states that are expressed within a larger environment \cite{Shafeeentropy}.

The process of quantum measurement allows for a state of a measurable ``quantum'' variable to be expressed by correlating each with a certain macroscopic state within the environment in the classical world.
The expression of one state onto another object thus is contingent upon the second object being able to interact with that first state. A different state should interact to yield a different type or degree of change in the object state due to certain interactions or because of the object being able to retrieve the effect of the state by being connected to a common environment that is effected by the state, if separate states are to be differentiated within the classical world.

Recently, the idea of variables and states in complex objects has been studied in detail \cite{Shafeesoc}. An object may be seen as an array of variables, each of which can exist in  different states.  The effect of the object onto another object is then a combined result of the variable-states of the first object interacting with variable-states of the second object \cite{Shafeesoc}.

For example, if the color of an object is a variable, and the variable is in state ``red'', the notion of this state is conveyed to an observing person when the light rays interacting with the first object also interact with the eye.

A state of a variable that does not interact with another variable-state within an object directly or via a medium is not comprehensible or expressed to the object.  If a variable interacts with another variable connected to an object or observer in a complex way, the direct existence of the first state may not be obvious to the object or the observer, but only some effects of a complex connection may be observed.

Again, if a variable is continuous, but the observer can interact with the variable only in a coarse grained manner, then the exact state of the variable may not be clear to the observer although the exact state would exist in the first object independent of the observer.  Hence the interaction interface and the compatibility of the interface with the leaking information also come into play when a variable state is expressed within another system. One such example of utilizing such a scheme would be the possibility of hidden variables in quantum mechanics assuming the spin is in reality determined for every angle in an electron, but the detector can only detect a +1 or a -1 \cite{BE1}.

\subsection{Types of Interactions Among Variables}

  The type of interactions among variable-states may depend on the variables themselves.  For example, gravitational interaction among matter depends on the value of the masses and varies continuously (as 1/$R^2$ and also as the direct product of the interacting masses).  For interaction among charges, similar charges (states) behave in a certain manner (repelling) while opposite charges interact differently.  We could analogously imagine interactions where specific states of a variable interact with certain states of another variable while remaining undetectable to other states.  Hence it is possible to imagine a spatially pervasive variable in a certain state that is only detectable at discrete spatial points in the macroscopic world, where the corresponding receiving or interacting variable-states are present.  Hence it is possible for spatially separated objects to be interacting in the same manner because of their interaction with that spatially pervasive common variable that is undetectable to others.  However, the time evolution of such variable states and the effects of interactions with such time evolutions due to interactions would be interesting to study.

\subsection{Interaction Among Levels}
Previously, we have discussed the effect of clustering in the expression of states in the classical world. In \cite{Shafeehier}, we have discussed the effects of having hierarchical structures in a complex system and have touched on the expression of variables at each level.  We have argued how at each level only a certain number of variables or parameters are expressed.  Similar arguments were put forth in terms of social networks, where an agent was modeled as an array of variables with possible states related to each variable \cite{Shafeesoc}.  The agents exhibited a finite number of states when allowed to interact with the environment and other agents because of constraints imposed on the interactions. These states were then allowed to evolve based on the interactions.  Sets of states were formulated to be energetically optimal in an interaction energy landscape, and some states were grouped as inhibitory pairs so that the interaction of one state with a certain variable inhibited the interaction with another state \cite{Shafeeperc} (Similar phenomena can be seen in case of certain enzymes binding to cell walls, when the binding of one type of enzyme inhibits another from binding).  These dynamic phenomena of expressing a group of changeable states were formulated by modeling such a complex identity as semi-closed \cite{Shafeehier}, and assuming that total interaction energy was limited to take into account mass, dimensions, charge etc physical characteristics and the laws of nature as constraints.

In the quantum measurement context, the quantum world may be seen as a level in a complex hierarchical structure, where the interface between the quantum and the classical world allow the leakage of a certain number of states (in the sense that the coupling of one state with another state at a certain hierarchy level prohibits the simultaneous coupling of an orthogonal state, hence prohibiting some states to be expressed when others are.) More examples of barring of the expression of simultaneous states within a shared object are given \cite{Shafeeperc}.

\section{Quantum Mechanics Derived from Hidden Variables}
Different models for explaining the results of quantum mechanics were mentioned previously.
The uneasiness about the concept of no particles existing at the quantum level, but having waves in superposed coexisting states that are later collapsed into particles, as proposed by the Copenhagen interpretation of quantum mechanics, had led to attempts of describing quantum measurements in terms of hidden variables.  In such theories, particles exist in the quantum world as well, and each of these particles also have specific defined states. However, the states depend on a hidden variable, or a set of hidden variables, often expressed with $\lambda$.  The statistical nature of $\lambda$ causes the measurements to yield different values. A way to look at hidden variable theories would be to simply take into account statistical variables associated with the measurement of a quantum variable so that the measurement process is similar to yielding one of the faces of a multidimensional dice in the Hilbert space when it is allowed to roll in the classical world that allows for the expression of fewer dimensions.  The question about whether quantum mechanics was complete or if hidden variable theories exist was raised by Einstein, Podolsky and Rosen \cite{epr}.
A local hidden variable theory assumes local reality as described below.  However, Bohm's model of hidden variables circumvents the necessity of the variable being local.

The Bohm picture and the Pilot Wave theory \cite{broglie,bohm} mentioned in the beginning of this paper where the position and momenta are hidden variables.  A guiding field obeying yields the same results as those found in quantum mechanics.

\subsection{Local Interaction and Propagation of Effects}
In most physical experiments expected to depend on specific variables, such as the detection of polarized light by a polarizer, the measurement processes leading to a result are considered to be local.  This ensures the possibility of reproducing the same result later \cite{wikiloophole}.
Although the entire classical world is seen as a connected system, the effects of interaction at a certain spatial and temporal point reaching another coordinate are restricted by the possible velocities of propagating or communicating the interaction, the maximum being the speed of light.  Hence, even if the state of any object in the universe depends on the history of the entire universe in a connected world, the effect of an interaction at a certain coordinate does not instantaneously effect another coordinate, but is subject to time lags and effects of dissipation and the intensities of interactions.

However, since the concept of the quantum world within the classical world depends on measuring quantum properties, results of experiments performed within the classical world are needed to discard any of the possible explanations of quantum measurement.

\section{Bell's Inequality}

Whether theories of hidden variables are plausible can be checked by a thought experiment proposed by Bell \cite{BE1}. The experiment assumes local realism, stating that variables are local, as is taken for granted in most experiments. The maximum speed for the propagation of information between two points is the speed of light.  If quantum potentials and guiding waves are related with particles as suggested in the Bohm/Pilot wave picture, the theory becomes non-local and hence is excluded from the set of local hidden variable theories tested by Bell's experiment.

 Bell's proposed experiment makes use of a pair of entangled electrons with a conserved total spin along a certain axis.  The electrons are then allowed to travel to two detectors, A and B, and measurements are carried out so that the result of measurement at the first detector is not known to the second detector when it measures the spin of the second electron.  When the detectors are aligned along the same axis but opposite direction, both of them show a 100 percent correlation in spin results based on initial conservation conditions. However, when the detectors are aligned at an angle with respect to each other, and the result of measurement from each detector is tallied up, quantum mechanics and local hidden variable theories exhibit different cut-offs in the upper bound of correlation.

  For local hidden variable theories, a statistically fluctuating common variable  $\lambda$ (which is allowed to be any general set of hidden variables), is associated with the measurement of a quantum variable.  Then the expectation value of measurements at a certain detector depends on the alignment of the detector and the hidden variable, $\lambda$.  One explanation for the correlation among spin measurement of the electrons by the two detectors is achieved by allowing the two electrons to carry the common hidden variable, which convey the information of spin conservation, to both detectors. So when each of the electrons is separately locally measured, the common hidden variable still preserves the correlation of spin among them.  Hence, if one electron collapses to either the up or the down state in one detector, the fate of the other electron's spin measured by the other detector is determined.

The original form of Bell's inequality was given by  \cite{bellin}
 \begin{equation}
 1+ C(b,c) \ge | C(a,b) - C(a,c)|
 \end{equation}

 where a,b and c are detector settings.

However, the form most used in conjunction with experiments is given by Clauser, Horne, Shimony and Hault \cite{CHSH}, which, for the case of hidden variables, gives the bound

\begin{equation}
C[A(a),B(b)]+C[A(a),B(b')]+C[A(a'),B(b)]-C[A(a'),B(b')] \le 2
\end{equation}

Here, C indicates correlation, A and B indicate expectations at detectors located at A and B, a, and a' are alternate detector settings at A, while b and b' are alternate detector settings at B.

By choosing specific angles for the detectors, this inequality can be shown to be violated if the measurements are related by an entangled pure wave.

So far experimental results favor quantum entangled waves, barring the possibility of local hidden variables except for the Bohm type models \cite{bellexp}.

\subsection{Superdeterminism}
One of the major views opposing Bell's experiment \cite{CHSreply} mentions the possibility of the detectors sharing the same hidden variable as that carried by the entangled pair of electrons.  The idea derives from the detectors and the source of the electrons being allowed to share a subset of common hidden variables in the past light cone.  This would invalidate Bell's assumption of local measurements being independent.

In his rebuttal against this argument \cite{BE2}, Bell mentions that such a situation would necessitate superdeterminism, requiring that every action in the universe is dependent on everything else, and hence it would also imply that the observers (experimenters) who are parts of the universe themselves do not have free will, and the angles of the detectors are also predetermined. This would invalidate the necessity of performing the experiment altogether.  A recent effort has included the construction of a continuous random field \cite{MORGAN} to explain possible common hidden variables in detectors and electrons.

\subsection{Questions about Bell's Inequality}

Some confusions may still prevail regarding this collapse of a joint entangled wave when one of the electrons in the pair is measured.  Since the two detectors are located reasonably far apart and the electrons are allowed to travel to the detectors from their common point of creation, after a certain time, the electrons are spatially adequately separated and their spatial wave functions can be considered disjoint  (The two electrons cannot be found at random spatial locations.  Their detection probabilities at a certain spatial point depends on the propagating wave packet with a specific momentum range that represents the electron.  Hence each electron has the probability of being detected by a detector only after a certain time.) As a result, two such electrons can be taken as distinguishable particles.  The collapse of one electron spatially does not cause the other to also come to a halt.  It is the internal degrees of freedom, namely the spins, that are entangled so that the measurement of one electron's spin causes the other to get fixed.
This phenomenon can raise the question whether the spins of the two electrons can be considered to be in a separate space (a subset of the Hilbert space) that does not involve the notion of spatial separation, where the collapse of one effects the other.
This idea of extended properties will be explored later.

\section{Our Aim}

Although quantum and classical worlds can be marked by the uncertainties related to the measurements of quantum objects {\it in} the classical world, where each object can be assigned a definite position and momentum, some confusions and ambiguities exist regarding what exactly the quantum world itself is, as was discussed in the previous sections.

In this paper, however, instead of trying to explain the entire quantum + classical world in a unified model, we concentrate in the interface of the two worlds, where the classical world is taken as a combination of clustered bodies with defined momenta and positions attributed to such bodies, and the quantum world is taken as microscopic objects exhibiting uncertainties in their momenta-position relations, as can be verified by experimentation performed on them in the classical world.

Whether the origin of such uncertainties is the actual existence of waves instead of particles in the quantum world, or a guiding wave function steering the particles according to Schrodinger's equation because of the uncertainties in momentum and position in measurement \cite{bohm, broglie} will not be crucial in the next part of the paper.

We start with assuming the detector ``sees'' superposed states (whether the states are actually superposed in the reality of a quantum world or whether they arise from the detector not being able measure the exact mometum/position and all other related variables).  We let the detector interact with the superposed quantum system in a specific way that leads to the expression of one of the states in the environment and the macroscopic world.

In the later part of the paper, we try to have some insight into the puzzling phenomenon of entanglement.  How the measurement of one quantum object by a detector leads to an entangled particle's fate being determined when measured by another detector that is separated from the first detector in such a manner that the information of the first collapse cannot reach the second detector by the time of the second measurement even at the speed of light, is examined. The confusing idea of the collapse of an entangled wave where the detection probabilities of the two electrons are spatially concentrated at two places (even as waves), but the measurement of the spin of one of the electrons causes the wavefunction of the second electron's spin to collapse in an entangled manner is examined.  A mechanism for the detectors acting on a class of common information that does not include superdeterminism \cite{BE2} is suggested.

\part{Review of Our Model of Quantum Measurement}

We first review a model of quantum measurement proposed by us \cite{Shafeewalk} where an image of the quantum system is formed within the detector, and then the reduction of states takes place by first passage random walk in the Hilbert space involving image-system state pairs. Each interacting system-image pair can be viewed as a team of betting partners against other such couples in a game. When the coefficient of one such pair becomes zero in the process of first passage random walk, which is analogous to going bankrupt, forcing one to leave the game, irreversible dimensional reduction takes place.  The process of loss of coherence takes place concurrently, as the unfavorable states among pairs also interact as spectators, appearing as cross terms that adjust their coefficients as the player pairs update their coefficients, until one state goes bankrupt, resulting in an irreversible zero coefficient for all terms involving interactions with that state.

The concepts of interactions among variables, selective interactions, states expressed in different hierarchical structures by means of interactions discussed in the previous sections are used in this model.

In the next section, we outline our proposed model for quantum measurement in the case of a two state quantum system. As mentioned before, it is irrelevant for this part of the paper, whether the results of quantum mechanics are based on the collapse of waves or on hidden variables.  We simply assume that the detector sees two states initially, and then interacts selectively within a hierarchical structure so that one of the two states is eventually reduced.

\section{Interactions Among Variables and Expression of Properties in Our Model}

In a previous paper \cite{Shafeewalk}, we have proposed a model for quantum measurement based on selective coupling among specific variable states, which are quantum states and their corresponding meso states created within the detector.

A meso or intermediate macro system is formed within the detector at the presence of the quantum system before one of the coupled quantum states is amplified to indicate a specific detector state within the macro world.  This meso system represents an image of the quantum system in the detector.

The process of image creation is analogous to the formation of the image of a charge placed in front of a conductor on the conductor surface.  Image formation in the detector can be modeled as below:

In a hierarchical model where each level is connected to the next, gradual amplification of signal is required so that one level in the hierarchy can influence the next. The meso system itself, which is the detector image, is conjured by an initial coupling of each quantum state with a corresponding state inside or near the detector.

 The initial coupling causes larger scale disturbances and cluster formation because of interactions among the many microscopic entities inside the detector, ensuring the initial disturbance can form a larger image, capable of acting as a buffer between the quantum system and the macroscopic detector.  Since these interactions may include quantum and classical terms, the resulting image is mesoscopic.  This image corresponds to the initial quantum system because the surface areas or the values of a specific resulting macroscopic property of the subclusters of microscopic systems attached to each of the states of the quantum system can be proportional to the coefficients of corresponding quantum amplitudes of the state because of the internal couplings and interaction mechanisms.

 Hence the amplitudes of the image can be represented by the coefficients attributed to separated components of a larger detectors meso-system, each  connected to a specific quantum state \cite{Shafeewalk}.

 The coupling between the image and the quantum system creates a joint system:

\begin{eqnarray}  \label{eq5}
|\psi\rangle_{SD} = \sum_i |a_i|^2 |i\rangle_S |i^*\rangle_D
+\sum_{i\neq j}a_i a_j^* |i\rangle_S |j^*\rangle_D
\end{eqnarray}

Here $|S>$ represents the quantum system state and $|D>$ is the detector state.

The joint function has terms indicating interaction/coupling between a quantum state and the corresponding detector state.  It also has cross terms that can be seen as passive spectators resulting from unfavorable interactions between a quantum state and a detector state corresponding to another quantum state.  These spectator cross-terms do not take part in the walk process directly, but adjust their coefficients as each of the states in a favorable pair $|a_i|^2 |i\rangle_S |i^*\rangle_D$ adjusts its coefficient.

Because of the coupling between components with the same coefficients (one from the quantum system and one from the image), terms corresponding to interaction between a quantum state and its corresponding image states have coefficients that are the squares of the initial quantum amplitudes.

 The coupled system is then allowed to perform a random walk within the Hilbert state. As the pairs undergo the walk, their coefficients change, and the spectator cross terms also readjust accordingly.  A decrease in one coefficient would imply parts of that pair leaking outside the detector, into the larger environment.

When one state is dissipated completely out of the detector, it becomes permanently zero (first passage random walk), and when all but one states (in their coupled detector-system pair form) leak out to the environment, the detector is locked to the remaining state, which is then amplified to give the detector state corresponding to the remaining quantum state.  Thresholds involved with the process of final amplification to a specific macroscopic detector state, as dictated by internal couplings, may necessitate the need for all but one pairs to leave the game so the winner takes all.

The leakage of all pairs except the final state to the environment in the scheme allows us to understand the disappearance of orthogonal terms described in the Zurek model \cite{ZUREK}. It also preserves the total angular momentum of the detector-environment system.  The notion of states in a hierarchical environment, where the detector is taken to be an object that can express one state as a whole explains why one of the possible quantum states is expressed in a macroscopic level and how the rest remain in the environment in a dispersed form, never expressing themselves as a state together.

As stated previously, the formation of an image and the existence of coupled pairs involving image and quantum states, each having the same amplitude, brings in terms that have coefficients that are squares of the amplitudes of the quantum wave.  The total amplitude squares add to one, which is the normalized summed probability of all possible expressed detector states.

Hence, the method of first passage random walks guarantees that the probability of any of the states remaining in the detector in the end depends on the initial coefficient of the coupled pair, which, as was modeled, is the square of the amplitude of the corresponding quantum state in the wave, as required by quantum mechanics.

\subsection{Description of the First Passage Random Walk}
In this section, we outline the walk for a two state system.
Let us say that before the process starts, the coupled detector-system pair is in the state
\begin{equation}  \label{eq7}
|SD_{in}\rangle = |a_0|^2 |00^*\rangle_{SD} + |a_1|^2
|11^*\rangle_{SD}+\quad \textrm{off-diagonal cross-terms}
\end{equation}

The process of reaching a final state can now follow a first passage random walk \cite{redner}.

In the two dimensional Hilbert space, where reduction is taking place, the initial point is taken to be $(x_0,y_0) = (|a_0|^2,
|a_1|^2)$. No restriction on the coordinates is imposed until the final stage of the walk process.

The relevant equations are:

\begin{equation} \label{eq8}
\frac{\partial c }{ \partial t}= D \frac{\partial^2 c}{\partial
x^2}
\end{equation}

Here the diffusion constant $D$ exhibits an effective strength
of interaction or a length related to the diffusion process, $t$ smooths out the small discrete steps of diffusion into a continuous scale and may be proportional to real time.

In Fig. 1 is a schematic representation.

This, however, is not a Feynman-type graph. 

\begin{figure}[ht!]
\begin{center}
\includegraphics[width=10cm]{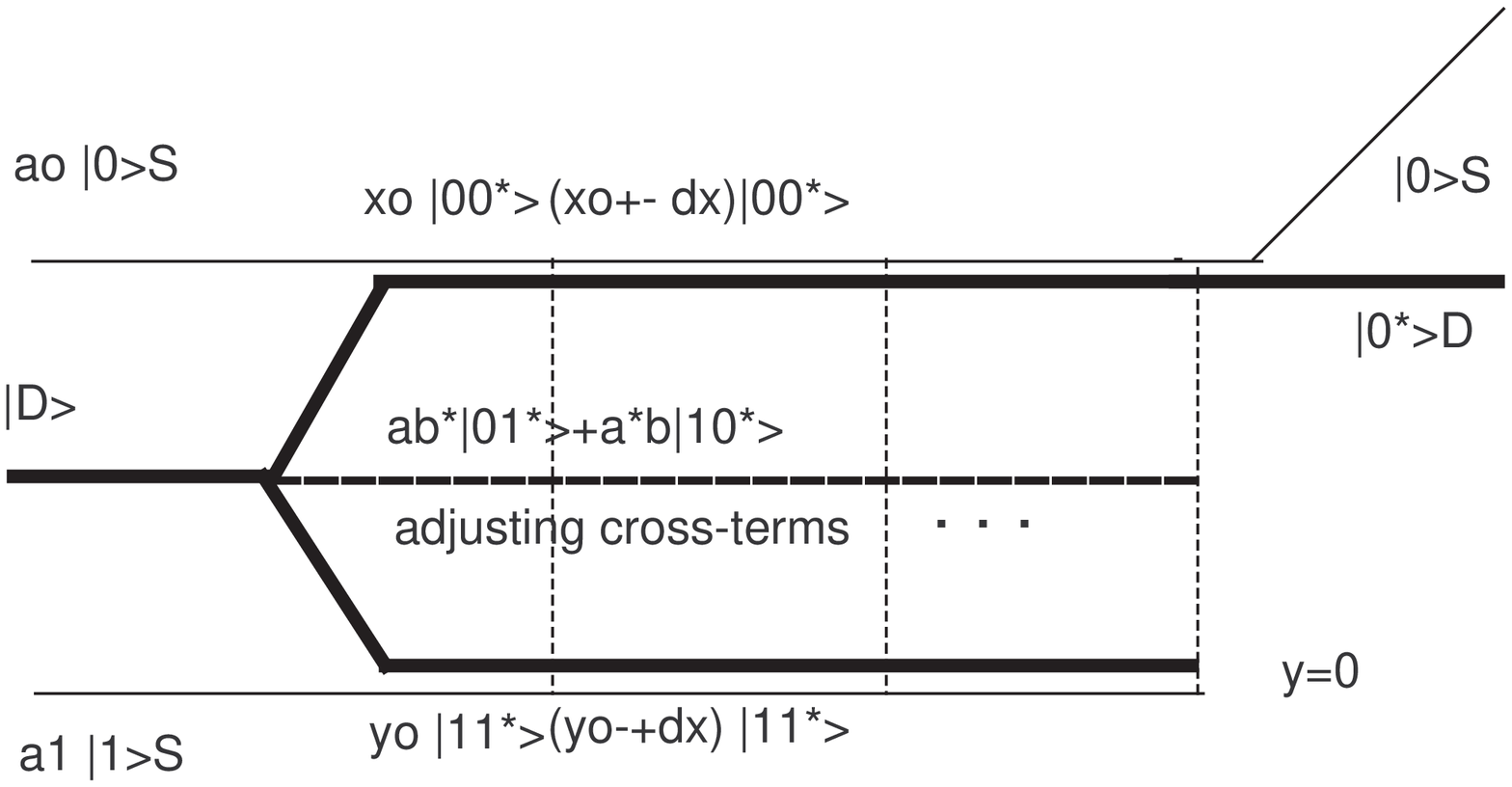}
\end{center}
\caption{\label{fig5025}Successive random transformations on competing eigenstates  $|00^*\rangle_{SD}$ and $|11^*\rangle_{SD}$ terminating
in the elimination of one.
}
\end{figure}

The sum over all possible step numbers can be equivalently represented by an integration over all $t$. It is more convenient to work
with the Laplace transform of $c$, which gives the transformed equations:

\begin{equation}  \label{eq9}
\frac{d^2\tilde{c}(x,s)}{dx^2}-(s/D)\tilde{c}(x,s)= -c(x,t=0)/D
\end{equation}

Here $s$ is the Laplace conjugate of $t$.

Initially, $c(x,0)$ is the delta function $\delta (x-x_0)$.

The boundary conditions are given by
$\tilde{c}=0$. These represent the absorbing walls $x=0,1$, where the first
passage walks stop.

The solution is given by the normalized Green's function:

\begin{equation}
\tilde{c}(x,s) = \frac{\sinh \left({\sqrt{(s/D)}}x_<\right) \sinh
\left({\sqrt{(s/D)}}
(1-x_>)\right)}{\sqrt{(sD)}\sinh\left({\sqrt{(s/D)}}\right)}
\end{equation}

Here $x_< = min(x,x_0)$ and $x_>= max(x,x_0)$.

The space derivatives with $s\rightarrow 0$ gives the probability of the process reaching an absorbing wall and hence an eigenstate:

\begin{eqnarray}
p(x=0) = D \left.\frac{ \partial \tilde{c}}{\partial
x}\right|_{s\rightarrow 0,x=0}= 1-x_0 = |a_1|^2   \nonumber\\
p(x=1) = -D \left. \frac{ \partial\tilde{c}}{\partial
x}\right|_{s\rightarrow 0,x=1} = x_0 = |a_0|^2
\end{eqnarray}

This is the same result expected by quantum mechanics.

 In the model outlined above, as was pointed out, interaction between states with the same coefficient within pairs that undergo a first passage random walk explains why the probability of observing a state behaves as the square of the amplitude. The need for decohered states to exist together in the environment before the reduction process begins is avoided, unlike in the method proposed by Omn\`{e}s, which also makes use of random walk \cite{omnes}.

Our model can be easily extended to \cite{Shafeewalk} a higher dimensional Hilbert space, where the walk stops upon reaching the vertex of a multidimensional complex, signifying reaching an end state that is expressed by the detector macroscopically.

\part{Bell's Inequality and Hidden Variables}
\section{Bell's Inequality and Hidden Variables}
In the first part, we have discussed the basic setup for Bell's thought experiment.  The experiment assumes local measurements of each in a pair of entangled electrons at two separated detectors.  It was shown that if the pair of electrons carried the information about conservation of total spin as common hidden variables contained within them, local disjoint measurements yielded a bound for correlation among spins of the two electrons which is different from that yielded by the predictions of quantum mechanics.

Let us briefly recapitulate the derivation of Bell's original 1964
form of the inequality to compare what will follow. The CHSH
\cite{CHSH} form is only a symmetrized form of the same and derivable
in an identical way.

Given the hidden variable $\lambda$ with the distribution
$\rho(\lambda)$ and given locality in the sense that the result of
any measurement for a given $\lambda$  at $A$ $(E^A =  \pm 1)$ and
at $B$ $(E^B = \pm 1)$ depend on only the orientations of the
measuring device at A and B and on $\lambda$, one gets the
expectation value for the spin correlation $C$ for different
orientations $b$ and $b'$ at B , but the same orientation $a$ at A:

\begin{eqnarray}\label{bell}
C(a,b)= \int d\lambda \rho(\lambda)E^A(a,\lambda) E^B(b,\lambda) \\
C(a,b)- C(a,b')  = \int d\lambda \rho(\lambda)E^A(a,\lambda)
E^B(b,\lambda)( 1+ E^B(b, \lambda) E(b', \lambda)
\end{eqnarray}

using the conservation law for the zero spin system

\begin{equation}
E^A(b, \lambda)+ E^B(b, \lambda) = 0
\end{equation}

with the resulting inequality

\begin{equation}
|C(a,b) - C(a,b')| \geq  1+ C(b,b')
\end{equation}

Since quantum mechanics gives

\begin{equation} \label{expect}
C(a,b)  = - \cos(a,b)
\end{equation}

the inequality can be violated by choosing appropriate directions
for a, b and b'.

However, if the detectors also were allowed to share the hidden variable, then the conditions for local measurements did not necessarily apply.  Bell had categorized this possibility as superdeterminism, which would necessitate the detector alignments also to be predetermined, making experiments with random orientations of the detectors impossible.

In the previous part, we have discussed a model for quantum measurement where variable-states of the system are allowed to interact and couple to corresponding variable-states within the detector.  It is the coupled states that perform a random walk in the Hilbert space. The idea of a walk in a Hilbert space raises the possibility of connection between spatially separated detectors in another variable space.
The proposition of the formation of an image of the quantum system within the detector brings to mind the possibility of the presence of a hidden variable in the detector that is dependent on the hidden variable shared by the two electrons in the form of an image.

In the next few sections, we first show how including hidden variables within each of the detectors locally, that also depended on the common shared variable $\lambda$ of the two electrons, may actually reproduce the quantum correlation. Then we discuss two possibile methods for introducing this $\lambda$ dependent hidden variable within the two detectors that do not necessitate superdeterminism.

\section{Detector Local Hidden Variables and Quantum Results}

In this section, we first mathematically derive how making a  detector- hidden variable depend on the electron hidden variable $\lambda$ can produce quantum correlations in Bell's experiment.

 Let us now first use unit vectors for the frames of measurement ${\bf a, b, b'}$, and
also a unit vector representing the hidden variable $\bf{\lambda}$,
which is related to the source (e.g. the decaying spinless
particle), which is shared by both A and B as it is carried by the
decay products reaching A and B, as in Bell's hypothesis. However,
we may also introduce local hidden variables $\mu^A$  and $\mu^B$
associated with the detectors A and B, with their own probability
distributions $\rho^A(\bf{\lambda}, \mu^A, {\bf a})$ and
$\rho^B(\bf{\lambda}, \mu^B, {\bf b})$:

\begin{eqnarray}
C({\bf a}, {\bf b}) =\sum_{\mu^A,\mu^B} \int d \bf{\lambda}
\rho(\bf{\lambda}) \rho^A(\bf{\lambda}, \mu^A, {\bf a})
E^A(\bf{\lambda}, \mu^A, {\bf a}) \nonumber \\
{       }  \rho^B(\bf{\lambda}, \mu^B, {\bf b}) E^B(\bf{\lambda},
\mu^B, {\bf b})
\end{eqnarray}

The integration over $\bf{\lambda}$ is over the angular variables
only, as we take it to be of unit length without losing generality.
The the influence of the local hidden variables $\mu^A$ and $\mu^B$
on the outcome of the measurement $E^A$  and $E^B$ in individual
events are deterministic in our picture, but also take into account
the values of $\bf{\lambda}$ and also the directions of the frames
${\bf a}$ and ${\bf b}$. As we shall see these variables may not
have any influence at all, or may also control the detected values
completely.

We notice that in this form, where locality is not violated, the
products $\rho^A E^A$ and $\rho^B E^B$ are no longer maneuverable as
in the Bell case using the simple relation $E^A( \lambda, a).E^A(
\lambda, a)= 1$, irrespective of $\lambda$ and of $a$ etc., used to
get Eqn. \ref{bell}. Hence Bell inequality does not follow as a
consequence of the local hidden variable's effect on the
determination of local measurements.

\subsection{A Specific Local Model with Quantum Expectation Values}

The general considerations of the previous section can be made
concrete with specific choices of the distribution functions for the
hidden variables and if they can reproduce the quantum mechanical
expectation value given in Eqn. \ref{expect}, then Bell's inequality
is automatically violated by the chosen model, indicating that
locality in this general form does not necessarily lead to Bell's
result.

We define the model by the relations (all the vectors are unit
vectors)

\begin{eqnarray}
\rho(\bf{\lambda}) = 1  \\
\rho^A(\bf{\lambda}, \mu^A=0, {\bf a})  = c_1 |({\bf a}. \bf{\lambda})|\\
 \rho^B(\bf{\lambda}, \mu^B=0, {\bf b})  = c_1 |({\bf
b}.\bf{\lambda})|\\
 \rho^A(\bf{\lambda}, \mu^A= \pm 1, {\bf a})  = c_2\\
 \rho^B(\bf{\lambda}, \mu^B=\pm 1, {\bf b})  = c_2\\
E^A(\bf{\lambda}, \mu^A=0, {\bf a})= sign ({\bf a}. \bf{\lambda})\\
E^B(\bf{\lambda}, \mu^B=0, {\bf b})= sign ({\bf b}. \bf{\lambda})\\
E^A(\bf{\lambda}, \mu^A=\pm 1, {\bf a})= \pm 1\\
E^B(\bf{\lambda}, \mu^B= \pm 1, {\bf b})= \pm 1
\end{eqnarray}

Since the contributions of $\mu^A, \mu^B = \pm 1$ cancels out in the
expectation value, the actual contribution to $C({\bf a}, {\bf b})$
comes only from the $\mu^A, \mu^B = 0$ part of the hidden parameter
space

\begin{equation}
C({\bf a}, {\bf b})  = c_1^2 \int d\bf{\lambda}({\bf a}.
\bf{\lambda})({\bf b}. \bf{\lambda})
\end{equation}

and with

\begin{equation}
c_1 = \sqrt{\frac{3}{(4 \pi)}}
\end{equation}

and $c_2$ given by the normlization condition

\begin{equation}
\int d\bf{\lambda}[ 4 c_2^2+ 4 c_1 c_2  |{\bf a}. \bf{\lambda}|+
 c_1^2 |{\bf a}. \bf{\lambda}||{\bf b}. \bf{\lambda}|]=1
\end{equation}

it gives

\begin{equation}
C({\bf a}, {\bf b}) = {\bf a} . {\bf b} = \cos(\theta_{ab})
\end{equation}

as quantum mechanics predicts. The inclusion of the $\mu^A, \mu^B=
\pm 1$ segments of the parameter space is required to normalize the
total probability, but it can be easily seen that they vanish when
${\bf a}$  and ${\bf b}$ are parallel or antiparallel. The overall
sign in the correlation is a matter of convention and can be
adjusted by taking the appropriate sign of the constants.

\section{Non-Spatial Coupling?}
The possibility of extended detector systems taking part in the measurement process has also been suggested elsewhere \cite{Jarisk}.

A way to view the non local nature required for any type of quantum prediction related to Bell's experiment would be to envision state spaces that are coupled to properties of specific objects but are not spatially dependent.  This would imply a mechanism for having shared common information among objects, or having extended state spaces circumventing the idea of spatial coordinates.  This notion does not imply superdeterminism because if the idea of space is taken away, the concept of spatially separated objects is also irrelevant, and two objects spatially far away may be very close in this new space.  It is also possible to couple specific objects or properties in this new space, which may appear in different spatial locations in the distance space. This would make it unnecessary to couple the entire environment to the variables represented in this new space, hence avoiding superdeterministic.

\subsection{Possible Problems Introduced by Delayed Measurements}

One question for such an extended Hilbert space scheme would be the handling of repeated measurements in an ensemble of correlated electron pairs.  Each pair of electrons created will have a statistically distributed $\lambda$ shared among both, which may be different from the $\lambda$ shared by another pair within the ensemble.  So if electron pair 1 has $\lambda_1$ then both the detectors are also expected to carry $\mu$ dependent on $\lambda_1$  and detector hidden variables corresponding to the measurements of electron pair 2 sharing $\lambda_2$ are expected to carry $\mu$ dependent on $\lambda_2$.
This is not a problem as long as the electrons are measured simultaneously.  However, if detector 2 is moved away in space so that when detector 2 makes the measurement, detector 1 has already measured electron 1, and another pair of electrons has also been created corresponding to $\lambda_2$, then the detection of electron 2 form the first pair corresponding to $\lambda_1$ may not be consistent because of the changed $\lambda$ within the detector.  In case of any set of distinguishable electrons, changing the time delay for measurement or having intermediate random electrons hit the detector before the detector records the second electron from the first pair may be problematic for this scheme.

A more comprehensive possibility of reduction of a joint $\lambda$ space where the detectors share a $\lambda$ space not dependent on their spatial separation may be worth looking into in more details.

\section{Possible Mirror Image Formation of $\lambda$}

  A second possibility may justify the equations given previously.  The scheme of the detector containing a hidden variable that depends on the electron's hidden variable $\lambda$ may be workable if the hidden variable distribution in the detector is caused by the electron's presence and hence the presence of the electron's hidden variable $\lambda$ itself.

This mechanism may actually be very similar to the one proposed in the last part, where a system variable interacts with a detector variable and evokes an image. If the presence of $\lambda$ in the electron creates a mirror image in the form of the distribution of the detector's hidden variable $\mu$, and the expectation value of the spin depends on both $\mu$ and $\lambda$, as was shown in the beginning of this part, it is possible to reproduce quantum correlations.

The interesting part of this scheme is that since the creation of the $\lambda$ dependent hidden variable in the detector depends on the electron carrying $\lambda$ being present in front of the detector, moving the detector, or bombarding the detector with stray electrons in between measurements do not effect the outcome. The problems with temporal delays disappear since the formation of the image depends on the presence of $\lambda$, and takes place when the electron and hence the electron's hidden variable interacts with the detector.

\section{Conclusion}
In this paper, we first reviewed some concepts regarding variables, interactions and expression of variables in hierarchical systems.  Then we reviewed a mechanism involving interactions among selective variables between a quantum system and a detector, producing a mirror image of the quantum object in the detector.
In the proposed scheme, the coupled pairs are then allowed to perform a first passage random walk in the Hilbert space so that one of the possible quantum states is expressed within the detector in the end, while the other states get dispersed within the environment. This model was shown to produce the predictions of quantum measurement. Lastly, we proposed a model based on a similar mirror image production within a detector in an attempt to shed light on the phenomenon of correlated local measurements of an entangled pair of electrons.  In future work, it might be interesting to examine in detail, the role of operators and the loss of unitarity when a transition is made from the quantum to the classical world.  The dynamics of Hamiltonians in a complex quantum wave, which itself is partly responsible for the Hamiltonian, and the role and emergence of classical interactions, giving rise to the arrow of time, might also be interesting to study in detail.

\section*{Acknowledgement} The author would like to thank Dr. George Jaroszkiewicz
for his comments on the original drafts on quantum measurement and Bell's inequality.  She would also like to thank Prof Geoffrey Sewell for reading the quantum measurement paper, Roland Omn\`{e}s for going over the draft for the measurement scheme in 2005, and to and Peter Morgan for sending his work on Bell's inequality.  She would like to express her appreciation to Lim Beng-Teck for his time reading the final draft as well.  Lastly, she would like to thank the editors of EJTP, especially Prof Ignazio Licata, for showing interest in her work.

\end{document}